\documentclass
[aps,prl,twocolumn,floatfix,english,showpacs,superscriptaddress,10pt]{revtex4-2}
\usepackage{graphicx}
\usepackage{booktabs}
\usepackage{amsmath}
\usepackage{physics}
\usepackage{amssymb}
\usepackage{colordvi}
\usepackage{verbatim}
\usepackage{xcolor}
\usepackage{mathrsfs}
\usepackage{epsfig}
\usepackage{lipsum}
\usepackage{amsfonts}
\usepackage{makecell}
\usepackage[unicode=true, breaklinks=false, pdfborder={0 0 1}, backref=false,
colorlinks=true, linkcolor=blue, urlcolor=blue, citecolor=blue]{hyperref}%
\providecommand{\U}[1]{\protect\rule{.1in}{.1in}}
\setcitestyle{numbers,square}

\begin{document}

\title{Giant enhancement of magnon transport by superconductor Meissner screening}

\author{Xi-Han Zhou}
\affiliation{School of Physics, Huazhong University of Science and Technology, Wuhan 430074, China}

\author{Xiyin Ye}
\affiliation{School of Physics, Huazhong University of Science and Technology, Wuhan 430074, China}

\author{Lihui Bai}
\affiliation{School of Physics, State Key Laboratory of Crystal Materials,
Shandong University, Jinan, 250100 China}

\author{Tao Yu}
\email{taoyuphy@hust.edu.cn}
\affiliation{School of Physics, Huazhong University of Science and Technology, Wuhan 430074, China}

\date{\today}

\begin{abstract}

Recent experiments observe the spin-wave-Meissner-current modes in ferromagnetic insulator-superconductor heterostructures, in which the coherently excited spin waves seemingly do not decay as usual beneath the superconductor strip [Borst \textit{et al}., Science \textbf{382}, 430 (2023)]. We interpret this phenomenon by demonstrating that the stray magnetic field emitted by the magnetization dynamics is reflected, focused, and \textit{enhanced} inside the ferromagnet by the supercurrent induced in the superconductor, such that the group velocity of spin waves is strongly enhanced.  Analytical and numerical calculations based on this model predict that the coherent transport of magnons is enhanced by close to $500\%$ for yttrium iron garnet capped by superconducting NbN with a decay length exceeding millimeters. Our finding may augment the performance of magnons in quantum information and quantum transport processing. 

\end{abstract}
\maketitle

\textit{Introduction}.---Spin waves or their quanta, magnons, are collective excitations of ordered magnetic moments, which can strongly hybridize with other degrees of freedom for classical information transduction, quantum information processing, and long-range pure-spin transport~\cite{Lenk,Chumak,Grundler,Demidov,Brataas,Barman,Yu_chirality,Cornelissen,Flebus}. In various magnonic devices, the intrinsic damping of spin waves is detrimental to their transport since it limits their possible propagation length, usually much smaller than photons and phonons. Ferrimagnetic insulator yttrium iron garnet (YIG) is currently the best
material for magnon transport due to its record low damping~\cite{travel1,travel2,travel3,Haiming}, long characteristic decay length of tens of micrometers~\cite{travel2,Haiming,Xiangyang}, and narrowest known linewidth 
of ferromagnetic resonance, resulting in a lifetime of a few hundred
nanoseconds~\cite{long_lifetime,Mingzhong}. The coherence time must be pushed to its limits to realize, implement, and assess the quantum states of magnons~\cite{Yu_chirality,Flebus,Q_inf1,Q_inf2,Q_inf3,Q_inf4,Q_inf5,Q_inf6}. Despite these merits, their dissipation parameters severely limit the appeal of employing magnons in quantum transport, which appears challenging to improve further via optimizing crystal quality.

Superconducting devices are instruments implemented to avoid Ohmic dissipation and employed on purpose in superconducting spintronics~\cite{Bergeret,study_1,study_2,yunyan_yao,RMP_2024}, quantum information~\cite{Q_inf1,Q_inf2,Q_information_2,Q_inf3,Q_inf4,Q_information_4,Q_inf5,Q_inf7}, and topological computation~\cite{topological}. 
Useful ferromagnet and superconductor heterostructure are pursued to transport spin information without dissipation and realize new functionalities~\cite{Bergeret,study_1,study_2,yunyan_yao,RMP_2024,Chumak_magnon_fluxon,Johnsen}. Superconductors with thicknesses comparable to London's penetration depth allow penetration of stray field emitted by spin waves, providing the electromotive force that drives the supercurrent and, in turn, causes the Ostered field that shifts the spin-wave frequency by many gigahertz~\cite{superconductor_gating_theory,similar_theory,superconductor_gate,superconductor_gate_1,superconductor_gate_2,superconductor_gate_3,Borst,Ghirri,Zhou,CPL_exp,PRA_exp,experiment,Gient_de,silaev}. However, the role of dissipationless supercurrent on magnon transport via this ``gating" effect has not yet been explored in previous theories~\cite{superconductor_gating_theory,similar_theory,superconductor_gate,superconductor_gate_1,superconductor_gate_2,superconductor_gate_3,Borst,Ghirri,Zhou,CPL_exp,PRA_exp,experiment,Gient_de,silaev}, raising the question that whether the supercurrent can ``shortcut" the magnon energy flow as in electronics.

Borst \textit{et al.} recently observed hybridized spin-wave-Meissner-current modes showing a modulated spin wavelength in the ferromagnetic insulator(FI)-superconductor (SC) heterostructures~\cite{Borst}. 
This experiment also finds that the coherently excited spin waves seemingly do not decay as usual beneath the superconductor strip with a significant enhancement of propagation length. It was speculated to be an artifact that an additional current may be induced in the SC strip when excited by the nearby microwave antenna~\cite{Borst}.  Less attention was, however, paid to the possible role of the supercurrent on the magnon transport in the FI-SC heterostructures.

\begin{figure}[htp]
\centering
\includegraphics[width=0.9\linewidth]{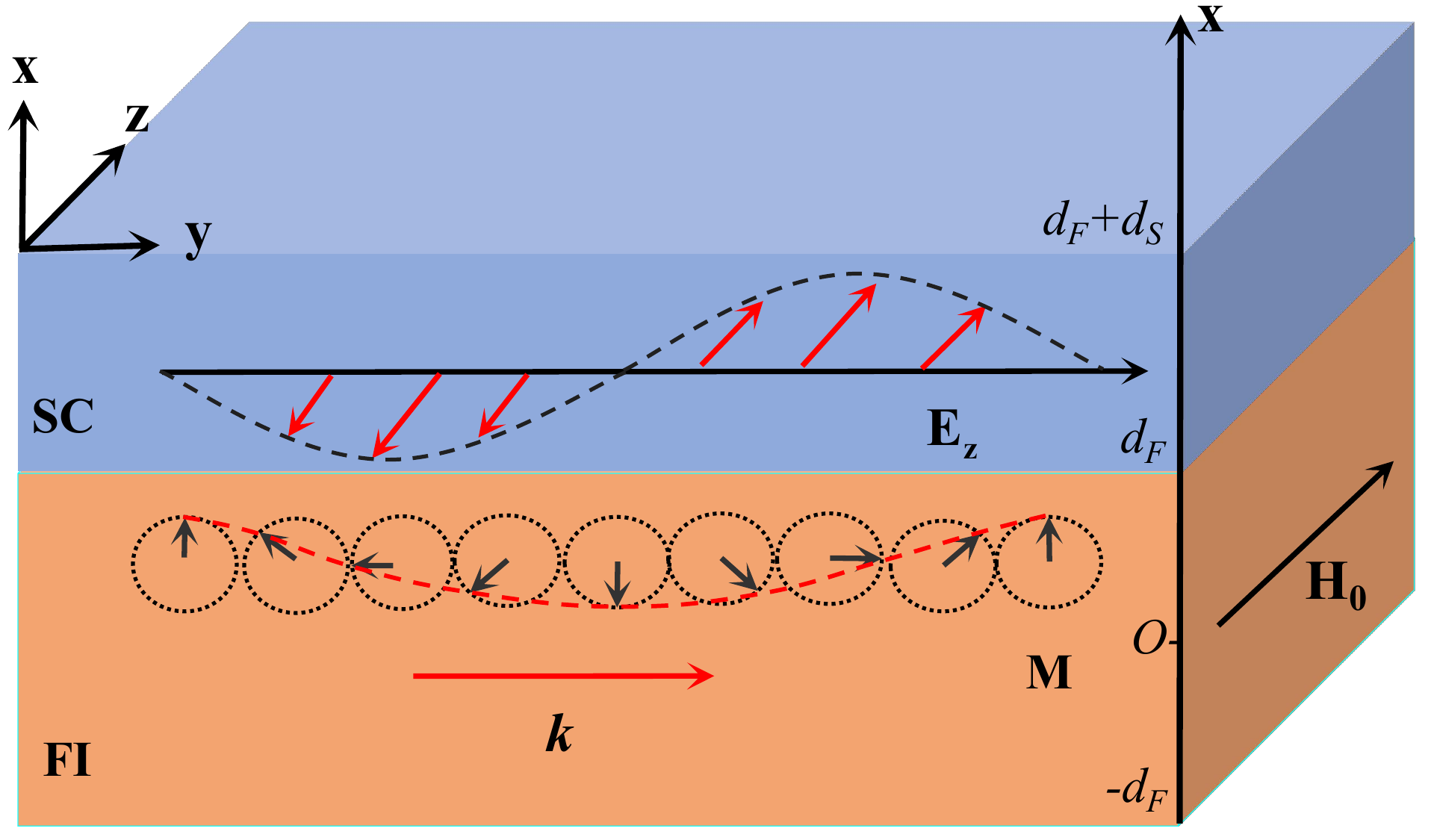}
\caption{Inductive coupling between magnetization dynamics and Cooper-pair supercurrent in FI-SC heterostructure. An in-plane external magnetic field ${\bf H}_0$ biases ${\bf M}_0$ to the $\hat{\bf z}$-direction. Spin waves propagating normally to the saturated magnetization ${\bf M}_0$ radiate the electric fields along ${\bf M}_0$. }
\label{model}
\end{figure}

In this Letter, we propose to use SCs for a local and unidirectional enhancement of magnon transport in thin FI by its inductive coupling to the Cooper-pair supercurrent as in Fig.~\ref{model}. 
We find that the electromotive force and supercurrent induced by the stray magnetic fields emitted by the magnetization dynamics can reflect, focus, and enhance the stray field inside the FI, such that the group velocity of spin waves is significantly enhanced while the net damping is not suppressed. We predict that by changing the temperature across the superconducting transition temperature (\textit{e.g.}, 5-10~K for NbN~\cite{NbN,NbN2,NbN3}), the magnon transport can be unidirectionally enhanced by up to $450\%$ for YIG film capped by superconducting NbN film, rendering a decay length exceeding millimeters, well beyond the present records~\cite{travel1,travel2,travel3,Haiming,Xiangyang}. Since quantum transport and quantum information are performed at cryogenic temperatures, our findings may augment the performance of magnons in these fields.

\textit{Reflection of magnetization radiation by proximity superconductors}.---Here we analytically calculate the electric and magnetic fields radiated by the magnetization fluctuation in the FI-SC heterostructure by solving Maxwell's equations and matching the boundary conditions at interfaces between the FI, SC, and vacuum. The exchange proximity effect between SCs and FIs can be intentionally suppressed by inserting an insulator spacer such as Al$_2$O$_3$~\cite{proximity1}, leaving the long-range dipolar interaction unaffected. As shown in Fig.~\ref{model}, a static in-plane magnetic field ${\bf H}_0$ orients the saturated magnetization ${\bf M}_0$ to the $\hat{\bf z}$-direction and the thickness of FI and SC are $2d_F$ and $d_S$, respectively. For the transverse magnetization fluctuation $M_{x,y}({\bf r}, t)={\cal M}_{x,y}(x)e^{i{\bf k}\cdot{\pmb \rho}}e^{-i\omega t}$ of frequency $\omega$ with amplitudes ${\cal M}_{x,y}$ and wave vector $\mathbf{k}=(0,k_y,k_z)$ propagating in the film plane ${\pmb \rho}=y\hat{\bf y}+z\hat{\bf z}$, ${\cal M}_{x,y}$ is nearly uniform across the film normal direction (i.e., the $\hat{\bf x}$-direction) when the film is sufficiently thin and the wavelength of spin waves ($\gg 2d_F$) is long~\cite{Borst,travel1,travel2,travel3,Haiming,Xiangyang}. 
Such magnetic fluctuation radiates the electric field ${\bf E}({\bf r},t)$ via the magnetization current ${\bf J}_M({\bf r},t)=\nabla\times{\bf M}({\bf r},t)$. According to Maxwell's equations~\cite{jackson}, the radiation obeys 
$\partial_x^2 \mathbf{E}({\bf r}, t)+(k^2_0-k^2) \mathbf{E}({\bf r}, t)
=-i\omega \mu_0  {\bf J}_M({\bf r},t)$,
 where $k_0=\omega\sqrt{\mu_0\varepsilon_0\varepsilon_r}$ with $\mu_0$, $\varepsilon_0$, and $\varepsilon_r$ being the vacuum permeability, permittivity, and relative dielectric constant, respectively. $k_0\sim 0.01~\mu {\rm m}^{-1}$ is tiny even for large $\varepsilon_r\sim 10$ for YIG and  $\omega\sim 100$~GHz.

The amplitudes $i{\cal M}_x=a_{\bf k} {\cal M}_y$ describe the ellipticity of spin waves in terms of $a_{\bf k}$.
$a_{\bf k}=1$ indicates the circular polarization, while the magnetic stray field renders  $a_{\bf k}\neq 1$. 
Analyzing the radiated electric fields by a single FI helps to understand the role of the capped SCs. We find the electric fields emitted by a single ferromagnetic film 
\begin{align}
{\bf E}^{(0)}({\bf r},t)={\pmb {\cal E}}^{(0)}(x)e^{i \mathbf{k} \cdot {\pmb \rho}-i\omega t}
    \label{single_E_z},
\end{align}
 governed by the amplitudes 
\begin{align}
{\cal E}^{(0)}_{x/y}&=+(-) \frac{  \omega \mu_0 {\cal M}_{y/x} k_z}{2 {\cal A}_k^2}\begin{cases}
\alpha_k(x), & x>d_F \\
-2+\delta_{+}(k,x), & |x| \leqslant d_F \\
-\alpha_k^{*}(x), & x<-d_F
\end{cases},\nonumber\\
 {\cal E}^{(0)}_z&=\frac{\omega \mu_0 {\cal M}_y }{2 {\cal A}_k} 
 \begin{cases}
 (\beta_{\bf k}-1)\alpha_k(x), &x>d_F\\
 (\delta_{+}(k,x)-2)\beta_{\bf k}+\delta_{-}(k,x),& |x| \leqslant d_F \\
 (\beta_{\bf k}+1)\alpha_k^{*}(x), &x<-d_F
 \end{cases},\nonumber
\end{align}
where ${\cal A}_k=\sqrt{k_0^2-k^2}\rightarrow ik$, $\alpha_k(x)=e^{i {\cal A}_k\left(x+d_F\right)}-e^{i {\cal A}_k\left(x-d_F\right)}$, $\delta_{\pm}(k,x)=e^{-i {\cal A}_k\left(x-d_F\right)}\pm e^{i {\cal A}_k\left(x+d_F\right)}$, and $\beta_{\bf k}=-ia_{\bf k} k_y/{\cal A}_k\rightarrow -a_{\bf k}k_y/k$.  When spin waves propagate normally to ${\bf M}_0$,  the electric field is linearly polarized along the saturated magnetization $\hat{\bf z}$-direction since ${\cal E}^{(0)}_{x/y}$ vanishes when $k_z=0$.  Particularly, for circularly polarized spin waves with $a_{\bf k}=1$, $\beta_{\bf k}\sim -a_{\bf k}k_y/k=-\text{sgn}{(k_y)}$ depends on the propagation direction:  above (below) the ferromagnetic film, the electric field exists only when $k_y>0$ ($k_y<0$), governed by the attenuation length $|k_y|^{-1}$, but vanishes when $k_y<0$ ($k_y>0$).

\begin{figure}[htp]
    \centering
    \includegraphics[width=\linewidth]{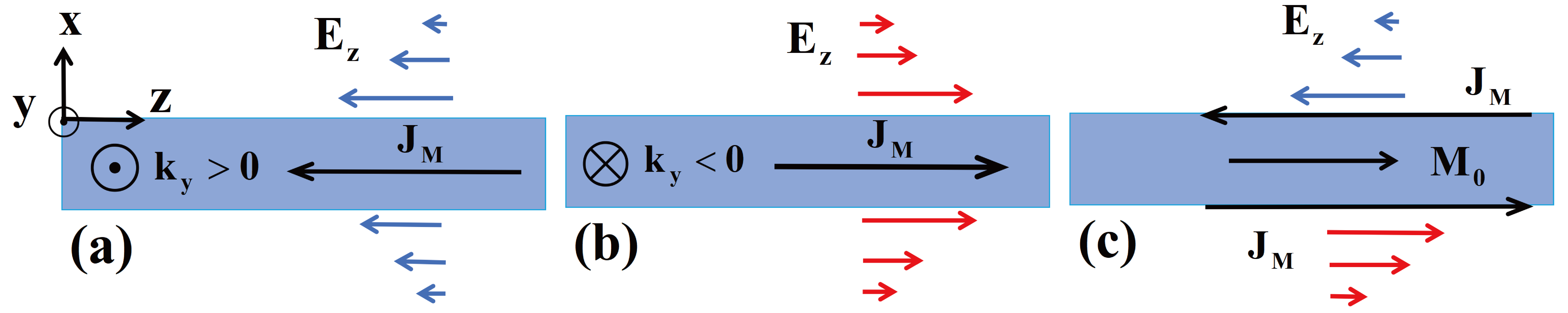}
    \caption{Snapshot of electric-field distribution radiated by the volume magnetization current when $k_y>0$ [(a)], when $k_y<0$ [(b)], and that by the surface magnetization current [(c)].}
    \label{Chirality_electric_field}
\end{figure}

This chiral electric field arises from the interference of surface and bulk contributions in the magnetization current ${\bf J}_M(x)=\nabla\times {\bf M}=[\delta(x+d_F)-\delta(x-d_F)-k_y]M_y\hat{\bf z}$. 
 The volume current $-k_y M_y \hat{\bf z}$ is proportional to the propagation direction $k_y$, which radiates the electric field $E_z$ decaying away from the surface. The direction $k_y$ determines the direction of $E_z$, as illustrated in Fig.~\ref{Chirality_electric_field}(a) and (b) for $k_y>0$ and $k_y<0$. On the other hand, the surface current $\delta(x+d_F)M_y-\delta(x-d_F)M_y$ is located at the upper and lower surfaces flowing in opposite directions, which radiates the electric field polarized in opposite directions above and below the magnetic film, as shown in Fig.~\ref{Chirality_electric_field}(c). Thereby, the net electric field, contributed by both the volume and surface magnetization currents, is suppressed (enhanced) below the film but is enhanced (suppressed) above the film when $k_y>0$ ($k_y<0$), exhibiting the unidirectionality.

We then demonstrate that the radiated electric (and magnetic) field by a single FI is strongly reflected, focused, and enhanced by the adjacent SCs. 
Inside the SCs, the electric field generates both normal current ${\bf J}_n$ and supercurrent ${\bf J}_s$~\cite{Tinkham,Janssonn}. The temperature $T$ affects them via the superfluid density $\rho_s(T)=\rho_e(1-(T/T_c)^4)$ and normal fluid density $\rho_n(T)=\rho_e(T/T_c)^4$ when $T<T_c$. $\rho_s+\rho_n$ equals the electron density $\rho_e$. Then the conductivity at low frequencies
\begin{align}
    \tilde{\sigma}(\omega)\approx\dfrac{\rho_n(T)e^2\tilde{\tau}}{m_e}+i\dfrac{\rho_s(T)e^2}{m_e\omega}
    = \sigma_n(T)+i\frac{1}{\omega\mu_0\lambda^2_L(T)},
    \nonumber
\end{align}
where $\tilde{\tau}$ is the relaxation time of electrons and $m_e$ is the electron mass, $\sigma_n(T)=\rho_n(T) e^2 \tilde{\tau}/m_e$ is the conductivity of normal fluid and $\lambda_L(T)=\sqrt{m_e/(\mu_0\rho_s(T)e^2)}$~\cite{penetration_depth} is London's penetration depth. The total electric current ${\bf J}=\tilde{\sigma}{\bf E}={\bf J}_n+{\bf J}_s$ screens the electric field via $\nabla^2 \mathbf{E}({\bf r}, t)+k_s^2\mathbf{E}({\bf r}, t)=0,$
where $k_{s}=\sqrt{\omega^2\mu_0\varepsilon_0-{1}/{\lambda_L(T)^2}+i\omega\mu_0\sigma_n(T)}$ is the electromagnetic wave number in the SC, which is complex governed by $\lambda_L(T)$ and $\sigma_n(
T)$. Although the normal fluid is included in the formalism, it plays a minor role when $T<0.95T_c$ since $\sigma_n(T)\propto(T/T_c)^4$ rapidly decays when decreasing the temperature,  referring to the Supplemental Material (SM) for details~\cite{supplement}.

The boundary condition governs the reflection and transmission of electric fields at interfaces. The in-plane ${\bf E}_\parallel({\bf r},t)$ is continuous at interfaces~\cite{jackson}. For the normal component $E_x$, there is no charge accumulation inside SCs at microwave frequency, such that $\nabla\cdot {\bf E}({\bf r},t)=0$ in it. On the other hand, $\partial_\parallel {\bf E}_{\parallel}({\bf r},t)$ is continuous, and thereby, combining with $\nabla\cdot {\bf E}({\bf r},t)=0$, $\partial_xE_x({\bf r},t)$ is continuous at interfaces. 
Finally, the current $J_x({\bf r},t) +\varepsilon_0 \partial_t E_x({\bf r},t)$ is continuous at interfaces due to the charge conservation. Carefully matching the boundary conditions go beyond the previous theoretical treatments~\cite{Borst,superconductor_gating_theory}, by which we obtain the magnetization radiation for all wave vectors $\bf k$, referring to the SM for details~\cite{supplement}.

Here, we focus on the Damon-Eshbach (DE) configuration with $k_z=0$ since these spin waves are modulated most strongly by SCs and were recently measured~\cite{Borst}. 
In this case, as that radiated by a single FI \eqref{single_E_z} only the electric field $E_z\hat{\bf z}\parallel {\bf M}_0$ exists, as addressed in Fig.~\ref{model}. The electric field emitted by a single FI \eqref{single_E_z} is reflected by the SC according to the wave-number dependent reflection coefficient 
${\cal R}_{k}=({\cal A}_k-{\cal B}_k)/({\cal A}_k+{\cal B}_k)$,
where ${\cal B}_k=\sqrt{k_s^2-k^2}$ and $k=|k_y|$ in the DE configuration, such that inside the FI 
\begin{align}
    E_z({\bf r},t)&=E_z^{(0)}({\bf r},t)\nonumber\\
    &+{\cal R}_kE_z^{(0)}({\bf r},t)|_{x=d_F}e^{-i{\cal A}_k(x-d_F)}.
    \label{electric_field_h}
\end{align}
For a large $k\gg \lambda_L^{-1}$, the electric field rapidly decays within the decay length $k^{-1}\ll \lambda_L$, such that $|{\cal A}_k|\sim |{\cal B}_k|$, implying vanished reflection ${\cal R}_k\sim ({\cal A}_k-{\cal A}_k)/(2{\cal A}_k) \rightarrow0$, i.e., the SC hardly affects the spin waves. On the other hand, when 
$k=0$, $|{\cal A}_k|\ll |{\cal B}_k|$ and ${\cal R}_k\sim -{\cal B}_k/{\cal B}_k=-1$, such that the electric field is totally reflected by the SC with a $\pi$ phase shift, resulting in no net supercurrents driven inside the SC, i.e., the magnon and Cooper-pair supercurrent decouples. However, the SC modulates the spin waves even for small $k$ since ${\cal R}_k$ rapidly departs from $-1$. The spin waves with $k\sim \lambda_L^{-1}$  are most strongly modulated by SCs.

The electric field \eqref{electric_field_h} induces the magnetic field ${\bf H}_d({\bf r},t)$ according to Faraday's Law $i\omega\mu_0[{\bf H}_d({\bf r},t)+{\bf M}({\bf r},t)]=\nabla\times {\bf E}({\bf r},t)$~\cite{jackson}. Inside the FI, the magnetic stray field ${\bf H}_d({\bf r},t)$ decays with the decay length $\sim 1/|k|$, while the attenuation length $\sim (k^2+1/\lambda_L^2)^{-1/2}$ in SCs is much shorter when $|k|<\lambda_L^{-1}$. Thereby, even for thin SCs of tens of nanometers, the decay is significant, while it is not equally important in the thin FI of similar thickness. We thereby take the spatial average for the stray field \textit{in the FI} along the normal $\hat{\bf x}$-direction:
\begin{align}
H_{d,y}&=M_y(
P_{k}e^{-kd_F}-1)\nonumber\\
 &+{\cal R}_{k}M_y{P_{k}e^{-kd_F}}\left(1-e^{-2kd_F}\right)\frac{1+a_{k_y}\text{sgn}{(k_y)}}{2} \nonumber\\
&\equiv \kappa_y(k_y)M_y,\nonumber\\
 H_{d,x}&=-M_x{P_ke^{-kd_F}}\nonumber\\
  &+{\cal R}_kM_x{P_ke^{-kd_F}}\left(1-e^{-2kd_F}\right)\frac{1+a_{k_y}^{-1}\text{sgn}{(k_y)}}{2}\nonumber\\
  &\equiv\kappa_x(k_y)M_x,
  \label{dipolar_field}
\end{align}
where the form factor $P_k=\sinh(kd_F)/(kd_F)$. 
As in the electric field \eqref{electric_field_h}, the dipolar field is also expressed as the one radiated by a single FI superposed by that reflected by the SCs. 
When $k_y= 0$
or ${\cal R}_k\rightarrow 0$, the influence of the SC vanishes, and the dipolar field recovers to that of a single FI, which, however, is \textit{enhanced} when ${\cal R}_k\rightarrow -1$ assuming $a_{k_y}\sim 1$. The dimensionless factors $\kappa_y(k_y)<0$ and $\kappa_x(k_y)<0$ are generally enhanced \textit{in magnitude} by the SC. The chirality of the dipolar field is now clear:  for circularly polarized spin waves with $a_{k_y}\rightarrow1$, $[1+a_{k_y}\text{sgn}(k_y)]/2\rightarrow 0$ and $[1+a^{-1}_{k_y}\text{sgn}(k_y)]/2\rightarrow 0$ when $k_y<0$, but becomes unity when $k_y>0$.

\begin{figure}[htp]
\centering
\includegraphics[width=1.0\linewidth]{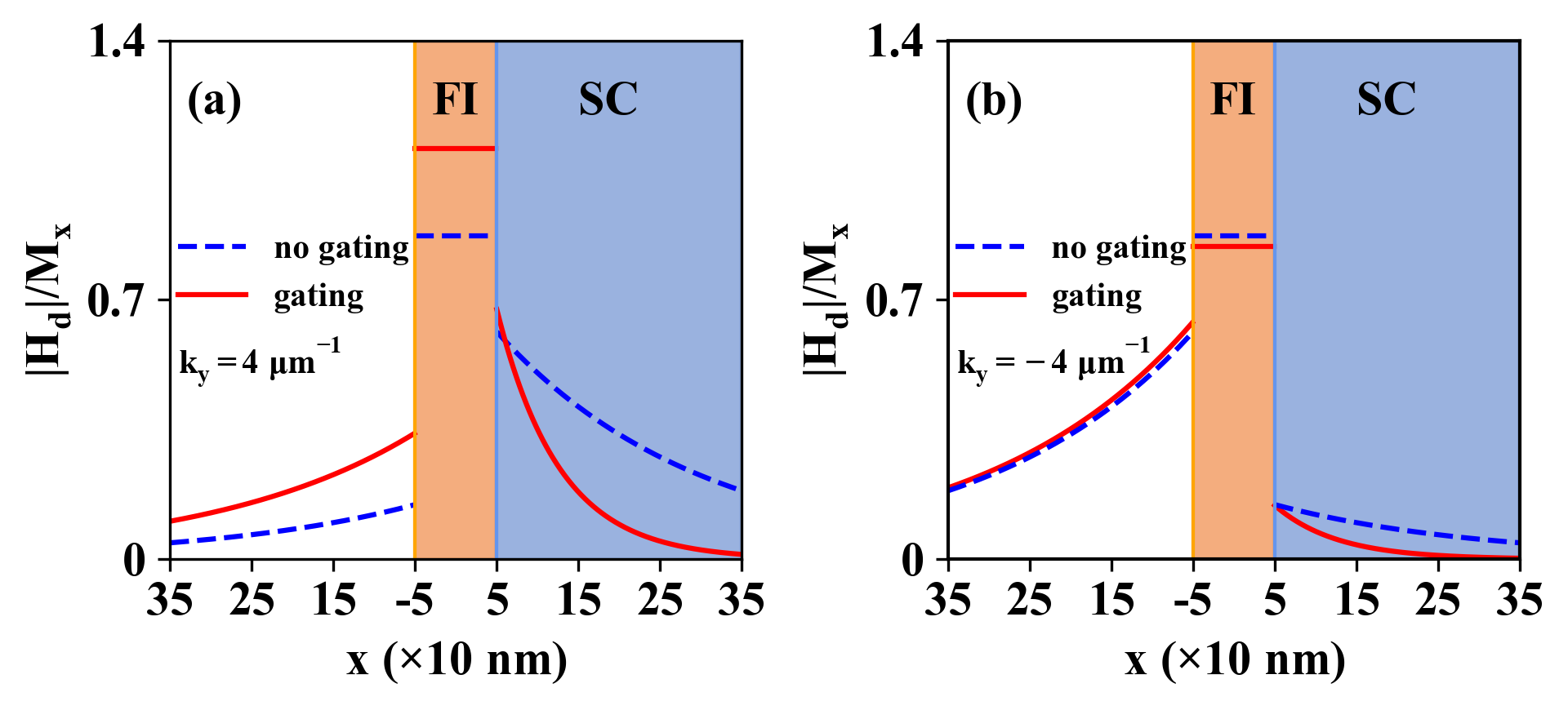}
\hspace{0.2cm}\includegraphics[width=0.96\linewidth]{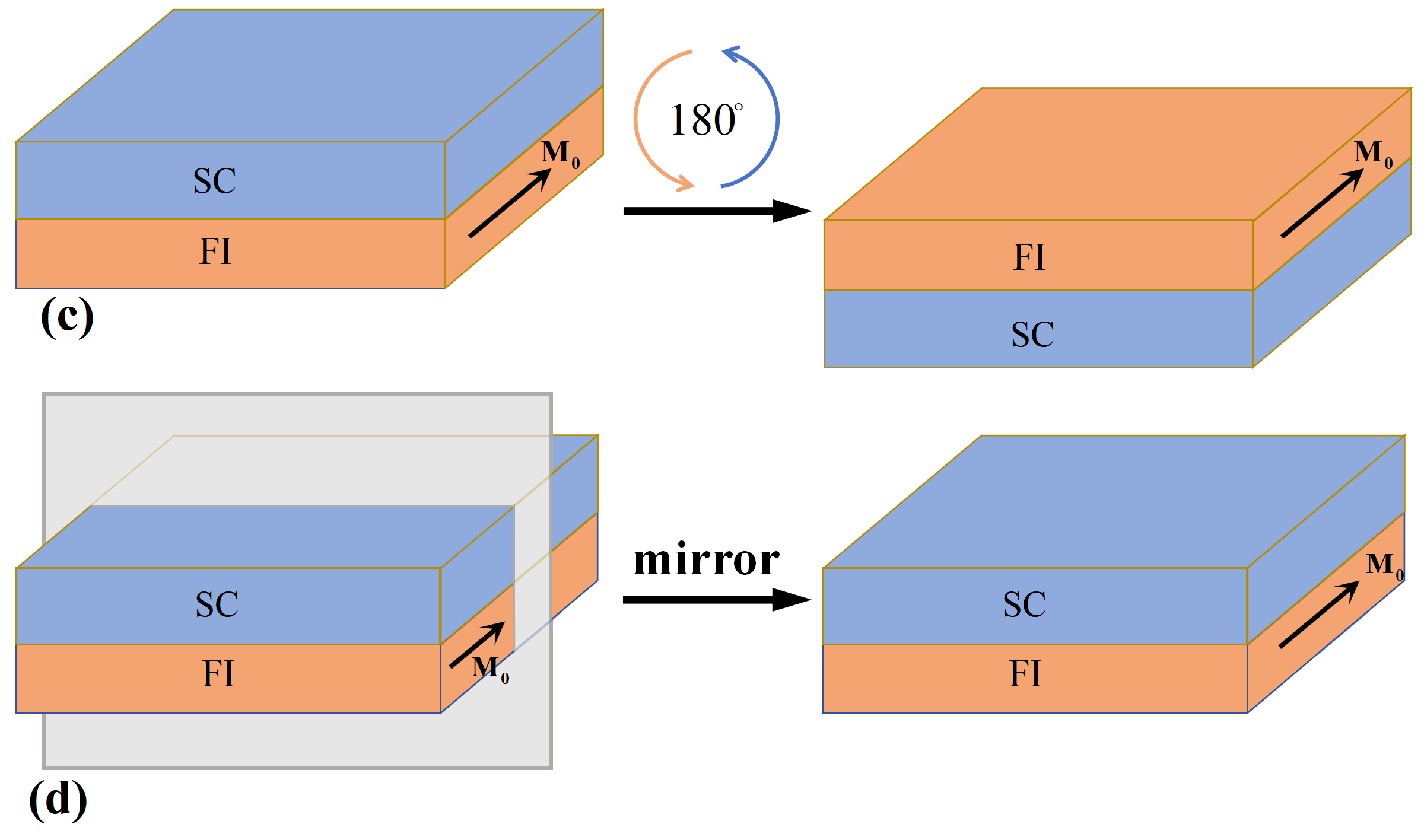}
\caption{Selective enhancement of radiated magnetic field by proximity SCs with  $k_y= 4$~$\mu {\rm m}^{-1}$ [(a)] and $k_y=-4$~$\mu {\rm m}^{-1}$ [(b)]. 
The blue dotted curve plots the magnetic field of an isolated FI with thickness $2d_F=100$~nm. The red curve illustrates the magnetic field of the FI-SC heterostructure with SC thickness $d_S=300$~nm and penetration depth $\lambda_L=80$~nm. (c) and (d) show the SC-FI heterostructure with twofold rotation symmetry breaking and mirror symmetry.}
\label{magnetic_field_dis}
\end{figure}

We now illustrate the effect of SCs on the stray field by comparing it with that of an isolated FI in Fig.~\ref{magnetic_field_dis}(a) and (b). We take $d_F=50$~nm, $d_S=300$~nm, and London's penetration depth $\lambda_L=80$~nm at $T=1.2~\text{K}\sim0.1T_c$ for NbN~\cite{NbN,NbN2,NbN3}. For the spin wave with $k_y=4$~$\mu {\rm m}^{-1}$, without gating its stray field mainly exists above the ferromagnetic film $x>0$ [Fig.~\ref{magnetic_field_dis}(a)]. When in proximity to the SC, the magnetic field at $x>0$ is reflected and decays faster inside the SC, which is then enhanced inside the ferromagnetic film. By contrast, when $k_y=-4$~$\mu {\rm m}^{-1}$, as shown in Fig.~\ref{magnetic_field_dis}(b), the emitted magnetic field without gating mainly exists below the ferromagnetic film $x<0$, thereby is hardly affected by the SC fabricated at $x>0$.

The selective enhancement of stray field is governed by the general symmetry principle for the system with translation invariance: as shown in Fig.~\ref{magnetic_field_dis}(c) and (d) the capped SC breaks the twofold-rotation symmetry around the saturated magnetization but retains the mirror symmetry with the mirror plane normal to the saturated magnetization.

\textit{Modulated dispersion, group velocity, lifetime, and ellipticity of spin waves}.---Such selectively enhanced dipolar field \eqref{dipolar_field} strongly affect the magnetization dynamics.  
In the Landau-Lifshitz-Gilbert (LLG) equation $\partial_t \mathbf{M}=-\mu_0 \gamma \mathbf{M} \times \mathbf{H}_{\rm eff}+\alpha_G{\bf M}/M_0\times \partial_t {\bf M}$, $\gamma$ is the gyromagnetic ratio, $\alpha_G$ is the Gilbert damping constant, and the effective magnetic field $\mathbf{H}_{\rm eff}$ includes the static field ${\bf H}_0=H_0\hat{\bf z}$, the dipolar field ${\bf H}_d$ (\ref{dipolar_field}), and the exchange field ${\bf H}_{\rm ex}=\alpha_{\rm ex}\nabla^2 {\bf M}$ with the exchange stiffness $\alpha_{\rm ex}$.
The damping term contributes to an extra imaginary component in the eigenfrequency, i.e., $\tilde{\omega}_{\bf k}=\omega_{\bf k}-i\Gamma_{\bf k}$. With $\{M_x,M_y\}\propto e^{-i\tilde{\omega}_{\bf k}t}$, we find the eigenfrequency, inverse lifetime, and ellipticity of the collective modes
\begin{subequations}
\begin{align}
    \omega({k_y})&=\mu_0 \gamma \sqrt{(H_{k_y}-\kappa_x({k_y})M_0)(H_{k_y}-\kappa_y({ k_y})M_0)},\label{dispersion}\\
    \Gamma({ k_y})&=\dfrac{\mu_0\gamma\alpha_G}{2}\left[2H_{k_y}-M_0\left(\kappa_x(k_y)+\kappa_y(k_y)\right)\right],\label{inverse_lifetime}\\
    a({k}_y)&=\sqrt{(H_{k_y}-\kappa_x(k_y)M_0)/(H_{k_y}-\kappa_y(k_y)M_0)},
   \label{ellipticity}
\end{align}
\end{subequations}
where $H_{k_y}=H_0+\alpha_{\rm ex}k_y^2M_0$. Equation~\eqref{ellipticity} provides the self-consistent equation for solving the ellipticity, with which we numerically find the dispersion \eqref{dispersion} and inverse lifetime \eqref{inverse_lifetime}. When $k_y>0$, $\kappa_{x}(k_y)<0$ and $\kappa_{y}(k_y)<0$ are both enhanced in magnitude by the adjacent SC, such that both the frequency \eqref{dispersion} and the inverse lifetime \eqref{inverse_lifetime} are enhanced. It is then unclear whether the transport can be definitely enhanced.

\textit{Enhancement of magnon transport}.---We then turn to calculate the characteristic decay length of the gated spin waves when excited by an external Oersted magnetic field ${\bf h}({\bf r},t)$ of frequency $\Omega$, emitted from a long stripline along the saturated magnetization $\hat{\bf z}$-direction, similar to the configuration in the experiments~\cite{Haiming,Borst}. 
It couples with the transverse magnetization via the Zeeman interaction~\cite{Zeeman} 
$\hat{H}_{\text{int}}=-\mu_0\int {\bf M(r)}\cdot{\bf h}({\bf r},t) d{\bf r}$.
According to the linear response theory, the excited magnetization by the Fourier components $h_\beta(x,k_y,\Omega)=(1/L)\int dy e^{-i k_y y} h_\beta(y)$, where $L$ is the sample length, of such field reads
\begin{align}
M_{\alpha}(x,k_{y},\Omega)&=\mu_{0}(\gamma\hbar)^{2}\int_{-d_F}^{d_F}dx^{\prime}\chi_{\alpha\beta}(x,x^{\prime},k_{y},\Omega)\nonumber\\
&\times h_{\beta
}(x^{\prime},k_{y},\Omega),
\end{align}
where the spin susceptibility 
\begin{equation}
\chi_{\alpha\beta}(x,x^{\prime},\mathbf{k},\omega)=-\frac{2M_0}{\gamma\hbar^2
}{\cal M}_{\alpha}^{\mathbf{k}}(x){\cal M}_{\beta}^{\mathbf{k}\ast}(x^{\prime})\frac{1}{\Omega-\tilde{\omega}_{\mathbf{k}}}.
\end{equation}
Here the normalized amplitude of the modes~\cite{Walker,Yu_non_Hermitian} 
\[
    {\cal M}_y^{\bf k}=\dfrac{1}{2\sqrt{2d_F}}\dfrac{1}{\sqrt{a_{\bf k}}},~~~{\cal M}_x^{\bf k}=-\dfrac{i}{2\sqrt{2d_F}}{\sqrt{a_{\bf k}}}.
\]
After summation over the Fourier components, the excited magnetization in the real space 
\begin{align}
  M_{\alpha}(y,t)&=\sum_{k_y}e^{ik_yy}M_{\alpha}(k_y,t)
  =4iL\mu_0\gamma d_F M_0\nonumber\\
  &\times {\cal M}_{\alpha}^{k_\Omega}{\cal M}_{\beta}^{k_\Omega*}\dfrac{1}{v_{k_\Omega}}e^{i (k_\Omega y-\Omega t)}e^{-\tilde{k}_{_\Omega}y}h_\beta(k_\Omega),
\end{align}
where $\tilde{k}_y=k_{\Omega}+i\tilde{k}_{\Omega}$ is the positive root of $\tilde{\omega}(\tilde{ k}_y)=\Omega$, and $v_{\bf k}=\partial\omega({\bf k})/\partial {\bf k}$ is the group velocity of spin waves. $\lambda=1/\tilde{k}_{\Omega}$ governs the decay length of excited magnetization.

We find $\tilde{k}_{\Omega}$ as follows. From the linearized LLG equation 
\begin{align}
     -i\omega M_x+\mu_0\gamma H_{k_y} M_y=\mu_0\gamma M_0 \kappa_y(k_y)M_y+i\alpha_G\omega M_y,\nonumber\\
    i\omega M_y+\mu_0\gamma H_{k_y} M_x=\mu_0\gamma M_0 \kappa_x(k_y)M_x+i\alpha_G\omega M_x,
    \nonumber
\end{align}
the eigenfrequency $\tilde{\omega}(k_y)$ is governed by the characteristic equation 
    \begin{align}
    \tilde{\omega}^2(k_y)&=\left[ \mu_0\gamma(H_{k_y}-M_0 \kappa_y({ k_y}))-i\tilde{\omega}(k_y) \alpha_G\right]\nonumber\\
    &\times\left[\mu_0\gamma(H_{k_y}-M_0 \kappa_x({ k_y}))-i\tilde{\omega}(k_y) \alpha_G\right].
    \label{characteristic_equation}
    \end{align}
    $\tilde{\omega}(\tilde{k}_y)=\Omega$ leads to 
   \begin{align}
    \Omega^2&=\left[ \mu_0\gamma(H_{\tilde{k}_y}-M_0 \kappa_y({\tilde{ k}_y}))-i\Omega \alpha_G\right]\nonumber\\
    &\times\left[\mu_0\gamma(H_{\tilde{k}_y}-M_0 \kappa_x({ \tilde{k}_y}))-i\Omega \alpha_G\right]\nonumber\\
&=\omega^2(\tilde{k}_y)-2i\Omega \Gamma(\tilde{k}_y)-\Omega^2\alpha_G^2\nonumber\\
&\approx\omega^2(\tilde{k}_y)-2i\Omega \Gamma(\tilde{k}_y),
\label{LLG_omega}
\end{align}
where the quadratic term $\Omega^2\alpha_G^2$ is disregarded since $\alpha_G\ll 1$. 
The solution $\tilde{k}_y=k_\Omega+i\tilde{k}_{\Omega}$ is a complex number containing a small imaginary component $\tilde{k}_{\Omega}\ll k_{\Omega}$, such that  $\Omega\Gamma(\tilde{k}_y)\sim \Omega\Gamma({k}_\Omega)$, and we are allowed to  
expand $\omega(\tilde{k}_y)$ to the leading order as 
\begin{align}
\omega(\tilde{k}_y)=\omega(k_{\Omega })+v_{k_{\Omega }} i\tilde{k}_{\Omega }.
\label{omega_com_k}
\end{align}
Substituting Eq.~(\ref{omega_com_k}) into (\ref{LLG_omega}) and equating the real and imaginary parts of Eq.~(\ref{LLG_omega}) lead to 
$\Omega=\omega(k_\Omega)$ and ${\tilde{k}_{\Omega}}={\Gamma(k_\Omega)}/{v_{k_\Omega}}$.
So the characteristic decay length 
\begin{align}
\lambda=1/\tilde{k}_{\Omega}={v_{k_\Omega}}/\Gamma(k_\Omega)=v_{k_\Omega}\tau(k_\Omega)
\label{decay_length}
\end{align} 
is simply governed by the group velocity and the magnon lifetime $\tau(k)=1/\Gamma(k)$.

\begin{figure}[htbp]
\includegraphics[width=1\linewidth]{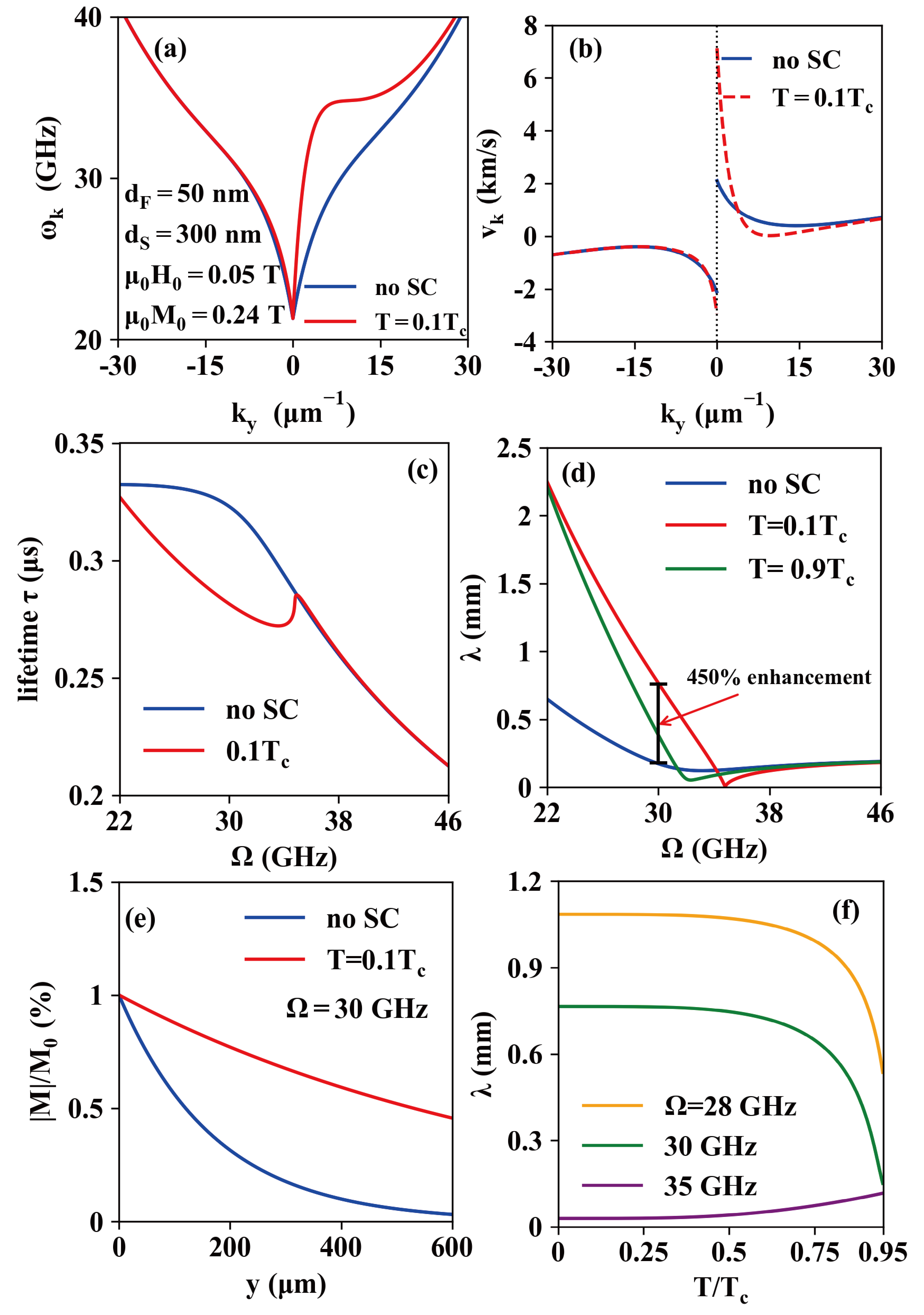} 
\caption{Enhancement of magnon transport in FI by an adjacent SC in the DE configuration ${\bf k}\perp{\bf M}_0$. (a), (b), and (c) plot the frequency shift, enhanced group velocity, and lifetime of magnons when the temperature $T=0.1T_c$ compared with that without SC gating. These modulations by the adjacent SC lead to a significant enhancement of magnon transport as demonstrated in the dependencies of the excited frequency $\Omega$ [(d)], characteristic decay length in the space [(e)],
 and temperatures [(f)]. }
\label{fig:dispersion}
\end{figure}

Figure~\ref{fig:dispersion} plots the modulated properties of spin waves by the adjacent SC in the DE configuration ${\bf k}\perp {\bf M}_0$.
In the calculation, we take the YIG film of thickness $2d_F=100$~nm, saturated magnetization $\mu_0M_0=0.24$~T~\cite{Borst,YIG_m0}, exchange stiffness $\alpha_{\rm ex}=3\times 10^{-16}~{\rm m}^2 $, and damping coefficient $\alpha_G=10^{-4}$~\cite{Mingzhong}, biased by the in-plane field $\mu_0H_0=0.05$~T. We use the superconducting NbN of thickness $d_S=300$~nm and London's penetration depth $\lambda_L=80$~nm at $T=0.1T_c=1.2$~K~\cite{NbN,NbN2,NbN3}. Comparing the dispersions when $T=0.1T_c$ and that in the absence of SCs, 
Fig.~\ref{fig:dispersion}(a) shows the magnon dispersion is strongly shifted upwards only when $k_y>0$. There is no frequency shift at the ferromagnetic resonance $k_y=0$ due to the total reflection of stray fields by the SC, and the shift vanishes when the wave vector is large since the exchange interaction dominates.
The frequency shift is then largest when $k_y\sim \lambda_L^{-1}$, resulting in the non-monotonic modulation of the group velocity as in Fig.~\ref{fig:dispersion}(b): the group velocity is strongly enhanced (slightly suppressed) when $k_y\in[0,\lambda_L^{-1}]$ ($k_y\gtrsim \lambda_L^{-1}$). Furthermore, the modulation on the magnon lifetime, which is slightly (not) suppressed when $k_y\in [0,\lambda_L^{-1}]$ ($k_y\gtrsim \lambda_L^{-1}$), is much weaker than that of the group velocity,
as shown in Fig.~\ref{fig:dispersion}(c).

According to Eq.~\eqref{decay_length}, magnon group velocity and lifetime jointly govern their characteristic decay length. As plotted in Fig.~\ref{fig:dispersion}(e), the decay of the excited magnetization becomes much slower due to the screening of the stray field by SCs. Figure~\ref{fig:dispersion}(d) and (f) plot the sharp dependencies of decay length on excited frequencies $\Omega$ and temperatures, in which we find the SC generally enhances the magnon transport when the excited wave vector $k_y\lesssim \lambda_L^{-1}$. Particularly, we estimate the enhancement for YIG can be as large as $450\%$, resulting in the decay length exceeding millimeters at lower frequencies.

\textit{Conclusion and discussion}.---In conclusion, we develop the model calculations of the inductive interaction between magnons and SC supercurrent and predict an efficient approach to unidirectionally enhance the transport of magnons in the microwave frequencies even in the best ferrimagnet YIG via the screening of its dynamical stray field by adjacent SCs. Such a mechanism is free of charge and spin-orbit-coupling, which is also suitable for metallic ferromagnets~\cite{metallic_damping} because of its relatively small conductivity. The frequency of some antiferromagnets~\cite{AFM_frequency1,AFM_frequency2} also lies in the control window, and the modulation may be efficient for both two polarized modes. At room temperature, the associated mechanism relies on the perfect diamagnetism of conventional SCs for microwaves, which may be replaceable by other materials with large diamagnetic susceptibilities~\cite{Pyrolytic_Graphite,Bismuth}. Our finding may stimulate explorations of magnon on-chip quantum information and transport in the context of superconducting magnonics.

\begin{acknowledgments}
This work is financially supported by the National Key Research and Development Program of China under Grant No.~2023YFA1406600, the National Natural Science Foundation of China under Grants No.~12374109 and No.~12274260, as well as the startup grant of Huazhong University of Science and Technology. We thank Toeno van der Sar, Gerrit Bauer, and  Yafei Ren for useful discussions.   
\end{acknowledgments}

\end{document}